\documentclass{iau} 
\usepackage{graphicx}

\def\gapprox{\;\rlap{\lower 2.5pt            
 \hbox{$\sim$}}\raise 1.5pt\hbox{$>$}\;}
\def\lapprox{\;\rlap{\lower 2.5pt            
 \hbox{$\sim$}}\raise 1.5pt\hbox{$<$}\;}

\title[High resolution polarimetry with AO systems] 
{Near-IR and Visual High Resolution \\ Polarimetric Imaging with AO Systems}

\author[H.M. Schmid]   
{Hans M. Schmid}

\affiliation{Institute for Particle Physics and Astrophysics, ETH Zurich, \\
  Wolfgang-Pauli-Str. 27, 8093 Zurich, Switzerland
} 

\pubyear{}
\volume{360}  
\jname{Astronomical Polarimetry 2020: New Era of Multiwavelength \\ Polarimetry}
\editors{H.Shinnaga, B-G Andersson, A.M.Magalh\~{a}es  \& E.Falgarone, eds.}
\begin{document}

\maketitle

\begin{abstract}
Many spectacular polarimetric images have been obtained
in recent years with adaptive optics (AO) instruments 
at large telescopes because they profit significantly
from the high spatial resolution. This paper
summarizes some basic principles for AO polarimetry, discusses
challenges and limitations of these systems, and describes
results which illustrate the performance of AO polarimeters
for the investigation of circumstellar disks, of dusty winds from
evolved stars, and for the search of reflecting extra-solar planets.
\end{abstract}

\firstsection 

\section{Science drivers for AO polarimetry}
Pushing observations to higher spatial resolution is a key for progress
for the investigation of the geometry and the physics of extended objects
in astronomy. The diffraction limited angular resolution of a
telescope is given by the ratio between wavelength and the telescope
diameter ${\cal{R}}\approx \lambda/D$.
High resolution ${\cal{R}}\lapprox 0.1''$ imaging polarimetry
in the visual to near-IR range requires therefore large telescopes
on the ground equipped with adaptive optics (AO) systems for the
correction of the atmospheric seeing or observations at
$\lambda< 1~\mu$m with the Hubble Space Telescope (HST).
This review focuses on the recent, significant
technical progress achieved with imaging polarimeters using
AO systems at large ground based telescopes.

\begin{figure}[t]
\begin{center}
  \includegraphics[width=13cm]{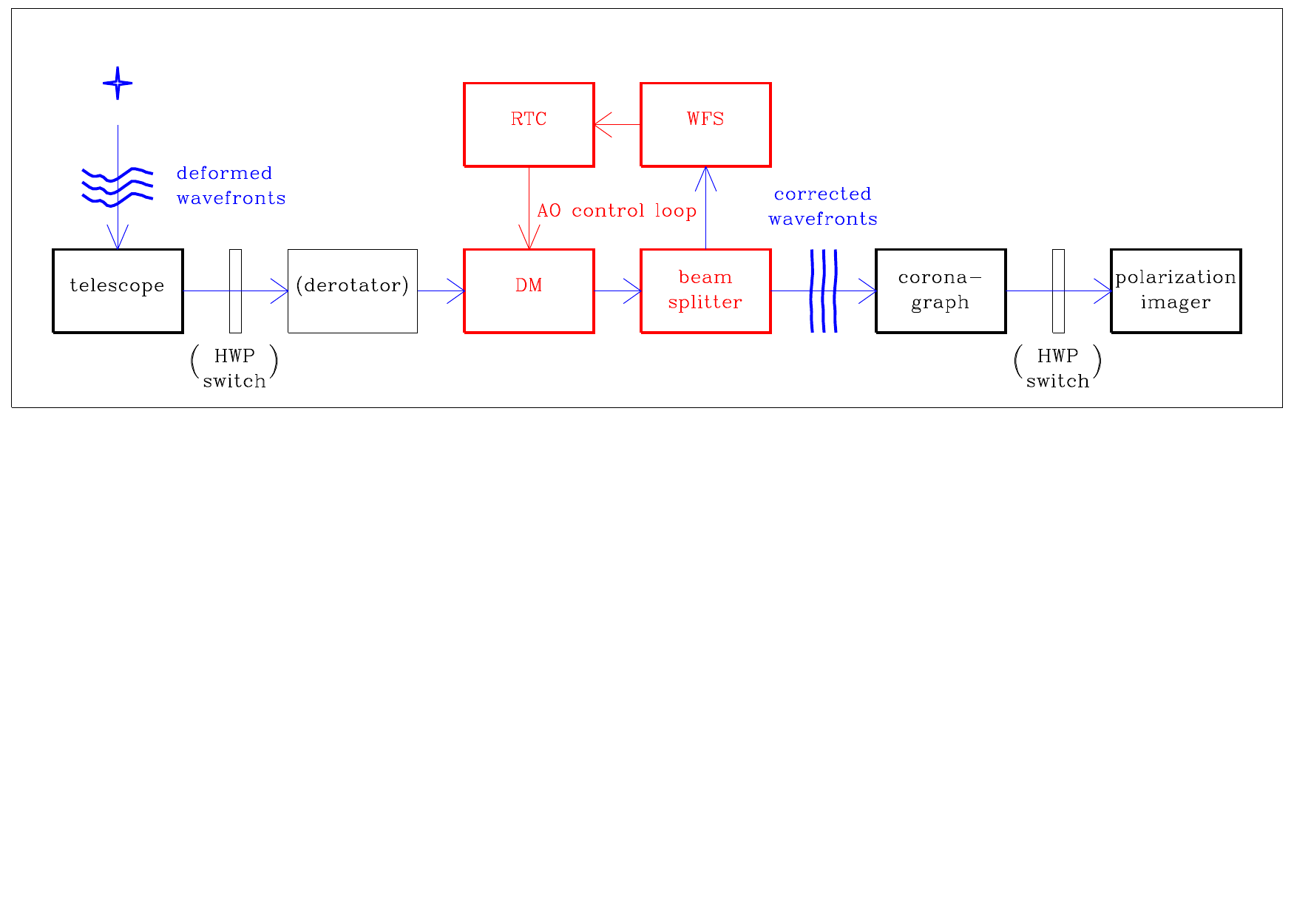}
\vspace*{-5.2 cm}
\caption{Light path and schematic block diagram for a telescope with
  AO system, coronagraph, and polarimetric imager. 
  Possible locations of a half wave plate (HWP) switch and
  a derotator are indicated (DM: deformable mirror, WFS: wave
  front sensor, RTC: real time computer).}
   \label{Figblock}
\end{center}
\end{figure}

The goal of AO systems is the correction of the wavefront
deformations introduced by the atmospheric turbulence with
a control loop (Fig.~\ref{Figblock}), which measures
the deviations with a wavefront sensor (WFS), calculates the corrections
with a real time computer (RTC), and applies them to a deformable
mirror (DM). This correction must be faster than
the turbulence time scale of a few milli-seconds and
this requires a bright photon source close to the science target.
The central stars of a circumstellar disk or shell, or a planetary
system are ideal wave front probes for good AO observations.

The AO system converts the seeing disk of a star into a point
spread function (PSF) with a diffraction limited core with
a width of $\approx {\cal{R}}$ and a residual halo with
about the size of the initial seeing disk
(e.g. \cite{Davis12}). For an 8~m telescope ${\cal{R}}$ is
$\approx 17$~mas in the R-band ($\lambda=0.65~\mu$m) and $\approx 42$~mas
in the H-band ($1.6~\mu$m).

The performance of an AO system is described by the
Strehl ratio $S$, which is roughly the fraction of light 
concentrated in the PSF core. $S$ depends for
a given system strongly on $\lambda$, for example VLT-SPHERE
reaches for a bright guide star under excellent
atmospheric conditions up to $S\approx 0.9$ in the H-band but
only $S\approx 0.6$ for the R-band (\cite{Fusco16}) and therefore
the PSF halo is stronger at shorter wavelengths.
The AO performance depends also on the seeing
and already average conditions can reduce $S$ by
30--50~\%. The Strehl ratio is also reduced for faint
guide stars (e.g. $>12^{\rm m}$) because of the lack of
photons for accurate wave front measurements (\cite{Milli17}).

The PSF core is typically so bright, that sensitive observations
of the stellar surroundings would heavily saturate the detector.
Many observations require therefore a 
stellar coronagraph (Fig.~\ref{Figblock}) to block the light
of the PSF core for $r\lapprox 0.1''$. For extreme AO systems,
like Gemini-GPI and VLT-SPHERE, the coronagraphs suppress also
the diffraction pattern of the PSF at 
separations $r\approx 0.1''$ to $0.4''$ (\cite{Sivarama01}).

The residual PSF halo is variable and typically stronger than the
signal of a faint companion, a disk, or the dusty stellar
wind. Differential techniques are required
to disentangle the light of a target from the halo of the
central star (e.g. \cite{Racine99}) and polarimetry is an ideal differential
technique to distinguish between the polarized scattered light of a
faint circumstellar target and the direct (and usually unpolarized) light
of the bright central star.
\smallskip

\noindent
The main science drivers for AO polarimetry are:
\begin{itemize}
\setlength{\itemindent}{0pt}
\item{} Bright stars with circumstellar scattering
  are ideal for a very good AO performance and often very interesting
  targets for polarimetric investigations.  
\item{} Separating and resolving circumstellar scattering regions
  from the star enhances significantly the measurable polarization signal. 
\item{} Polarimetry is an ideal differential technique for high
  contrast observations of polarized sources.
\item{} Often it is much easier to detect and image a target with differential
  polarimetry, while measurements of the intensity requires a follow-up
  in a second step.
\end{itemize}

\section{Technical challenges for polarimetry with AO systems}

\subsection{Historical developments}
The first AO systems with polarimetric modes were the 
Cassegrain instrument Subaru-CIAO (\cite{Murakawa04}) and the
Nasmyth instrument VLT-NACO (\cite{Lenzen03}), and they 
delivered very promising polarimetric result
for disks and circumstellar shells (\cite{Apai04,Murakawa05}). 
They also showed that simultaneous measurements of opposite
polarization modes $I_\perp$ and $I_\parallel$,
and the polarimetric self-calibration with a half wave plate (HWP) switch are
essential for good results, and that the calibration of systematic polarization
effects for the Nasmyth instrument NACO is difficult (\cite{Witzel11}).
This triggered efforts for better correction procedures for the 
polarization of the Nasmyth instruments VLT-NACO and
Subaru-HiCIAO and disk images with much higher quality were achieved
(e.g. \cite{Quanz11,Hashimoto11,Muto12}).
In addition, the lessons learned from the ``early'' AO polarimeters
were considered for the designs of the
subsequent extreme AO instruments Gemini-GPI (\cite{Macintosh14})
and VLT-SPHERE (\cite{Beuzit19}) which provide currently
the best performance in AO polarimetry.

Table~\ref{Systems} lists key system parameters and corresponding
references of widely used systems, which produced a large fraction
of the recent scientific result in  
AO polarimetry. The table illustrates the progress in AO performance from the
older Subaru-HiCIAO system (the parameters of VLT-NACO are similar),
to the newer extreme AO systems Gemini-GPI and VLT-SPHERE. Also of interest
are the very different polarimetric designs of the instruments,
which are described below.

\begin{table}
  \begin{center}
    \caption{Rough parameters for widely used AO polarimeters, which are available or were available
    until recently at 8m class telescopes.}
  \label{Systems}
 {\scriptsize
  \begin{tabular}{lcccc}\hline 
              &  ~~~{\bf Subaru}  & {\bf Gemini-South} & \multispan{2}{\hfil{\bf VLT-SPHERE}$^{1}$ \hfil}  \\ 
              &  ~~~{\bf HiCIAO}$^2$ & {\bf GPI}$^3$   & {\bf IRDIS}$^{4}$   & {\bf ZIMPOL}$^5$ \\
  \hline  
  AO system     &  AO188$^6$    & (GPI)         & \multispan{2}{\hfil SAXO$^7$ \hfil}    \\
~~~ in operation  & 2009-2015 & 2013-2020$^a$ & \multispan{2}{\hfil since 2015 \hfil}  \\
~~~ DM actuators  &  188           & $43\times 43$ & \multispan{2}{\hfil $41\times 41$  \hfil}    \\  
~~~ Strehl ratio$^b$
              &  $\approx 0.4$, H-band$^6$
                               & $\approx 0.9$, H-band$^{3}$
                                               & $\approx 0.9$, H-band$^{7,8}$ & $\approx 0.6$, R-band$^{7,8}$ \\
~~~ guide star limit$^{c}$ & $I<11^{\rm mag}$ & $I<10^{\rm mag}$ & $R<14^{\rm mag}$ & $R<11^{\rm mag}$ \\
~~~ coronagraphy  &  X         &  X            & X          & X           \\
\noalign{\smallskip}
polarimetric imager \\
~~~ polarimetry type & double beam & integral field pol.$^{9}$ & double beam$^{4}$ & fast modulation$^{5}$ \\
~~~ wavelength range &  $0.85-2.5~\mu$m & $0.9-2.3~\mu$m &  $1.0-2.3~\mu$m  & $0.52-0.9~\mu$m \\        
~~~ spatial resolution & $40-60$~mas    & $30-60$~mas    &  $30-60$~mas     & $20-25$~mas     \\ 
~~~ field of view &  $10''\times 20''$
                              & $2.7''\times 2.7''$
                                             & $11''\times 11''$
                                                    & $3.6''\times 3.6''$ \\
\noalign{\smallskip}
\multispan{2}{telescope and instrument polarization\hfil} \\
~~~ focal station &  Nasmyth       & Cassegrain    & \multispan{2}{\hfil Nasmyth \hfil} \\ 
~~~ HWP switch      & after AO   & after AO    & \multispan{2}{\hfil before derotator and AO \hfil} \\
~~~ telescope polarization
               & moderate    & small$^{10}$    & moderate$^{11}$   & large/compensated$^{5,12}$ \\
~~~ derotator cross talks
               & large       & none            & large$^{12}$   & large/controlled$^{12}$  \\         
 \hline
  \end{tabular}
  }
 \end{center}
\vspace{1mm}
 \scriptsize{
   {\it Notes:} 
   $^a$at Gemini-South, will be upgraded and mounted at Gemini-North in future,
   $^b$for a bright guide star and excellent atmospheric conditions, $^c$Strehl ratio
   degraded by a factor of $\approx 2$ because of guide star ``faintness'' \\
   {\it References:} 
   $^1$\cite{Beuzit19}, $^2$\cite{Hodapp08}, 
   $^3$\cite{Macintosh14}, $^4$\cite{Dohlen08},
   $^5$\cite{Schmid18}, $^6$\cite{Suzuki10}, $^7$\cite{Fusco16},
   $^{8}$\cite{Milli17},
   $^{9}$\cite{Perrin15}, $^{10}$\cite{Wiktorowicz14},
   $^{11}$\cite{deBoer20},
   $^{12}$\cite{Bazzon12}. 
   }
\end{table}

\subsection{Polarization imagers}
{\sl Double beam polarimeters} for linear polarization measurements
are based on a polarization beam splitter, often a Wollaston prism,
and a rotating HWP. This provides two images side by side on the detector with
``opposite'' polarization intensities $I_\perp$ and $I_\parallel$ from
which one can determine the differential signal $I_\perp$--$I_\parallel$.
The two images are taken simultaneously and this is essential for
the suppression of the strongly variable speckle noise
present in AO observations. The HWP is
used for the selection of the Stokes
$Q$=$I_0$--$I_{90}$ and $U$=$I_{45}$--$I_{135}$
parameters and as switch for the compensation of the instrument polarization
$q_{\rm inst}I$ and $u_{\rm inst}I$. A rotation of the HWP by $45^\circ$
switches the sign of the incoming $Q_{\rm in}$-signal
(or $U_{\rm in}$) for two consecutive measurements $Q^+$ and $Q^-$,
while the instrument polarization remains unchanged. The difference \\
\centerline{$Q^+ - Q^-=(Q_{\rm in}+ q_{\rm inst}I) - (-Q_{\rm in}+ q_{\rm inst}I) = 2\,Q_{\rm in}$} \\
\noindent
compensates then the instrumental polarization. The use of a HWP switch
is essential for all AO polarimeters. The basic double beam set-up was
used by the ``early'' AO polarimeters Subaru-CIAO and
VLT-NACO, but also for Subaru-HiCIAO. 
For SPHERE-IRDIS the principle is the same, but it uses a ``grey''
beam splitter plate and polarizers to avoid the differential
aberrations of a Wollaston prism between the two beams (\cite{Dohlen08}).
Double beam polarimeters provide a high efficiency and typically
a large field of view. 

\noindent
{\sl Integral field polarimeter:} This is an innovative concept
used as polarimetric mode in combination with the integral field
spectrograph (IFS) of Gemini-GPI which proved to work very
successfully (\cite{Perrin15}). The IFS
uses a 2D-array of micro-lenses in the focal plane,
producing spots which are then dispersed with a grism producing an array
of about 36\,000 low resolution spectra on the detector. In polarimetric
mode, the grism is replaced by a Wollaston prism producing an array of 
double spots with ``opposite'' polarizations $I_\perp$ and $I_\parallel$
instead of the low resolution spectra.
The field of view is given
by the focal plane sampling of the micro-lens array which
is smaller (for a given detector size) than for the ``standard''
double beam polarimeters.

\noindent
{\sl Fast modulation polarimeter:} SPHERE-ZIMPOL is based
on a fast polarization modulation concept using a liquid crystal
device and a polarizer, which converts the polarization modulation
into an intensity modulation between $I_\perp$ and $I_\parallel$.
A CCD-detector is used for the signal demodulation with the
main advantage that $I_\perp$ and $I_\parallel$ are registered with
the same detector pixels so that flat-fielding effects are minimized.
The modulation is faster than the speckle variations and $I_\perp$
and $I_\parallel$ are registered essentially
simultaneously (\cite{Schmid18}). The system is optimized
in various ways for polarimetry, achieves a high dynamic range and
a very high sensitivity of $10^{-5}$ for the search of reflecting
planets around bright nearby stars in the visual range.
Less good, when compared to double beam imagers, are the parameters
for the polarimetric efficiency (70--90~\%), the small
field of view, and the bright guide star limit for the AO system,
because ZIMPOL shares the light with the visual WFS. 

\subsection{Telescope and instrument polarization effects}
``Smart'' polarimeters have a simple design which minimizes
telescope and instrument polarization effects with a Cassegrain instrument
and a straight through beam. Subaru-CIAO (\cite{Murakawa04})
and Gemini-GPI (\cite{Perrin15}) followed as far as possible this principle and
therefore the polarimetric calibration of the data is relatively
simple (e.g. \cite{Wiktorowicz14}).

All other systems in Table~\ref{Systems} are ``not smart'', but complex,
large, and located at the Nasmyth focus because this is favorable
for many AO aspects. Therefore, the telescope and
instrument polarization effects are complicated and require a good
concept for the measurements of the polarization 
without compromising the performance of the AO system and the
coronagraph. A few typical problems are:
\begin{itemize}
\setlength{\itemindent}{0pt}  
\item{} the polarization effects of the inclined telescope mirror M3
  for Nasmyth systems,
\item{} ``rotating'' instrument polarization effects introduced by an image derotator,
\item{} static instrument polarization effects from inclined optical components.
\end{itemize}
\smallskip

\noindent
There are many ways how these problems can
be solved and each system must be considered individually.
The HWP switch discussed for the polarization imagers can be
placed early in the beam (Fig.~\ref{Figblock})
to compensate for the instrument polarization of many components. However,
this approach does not correct for polarization cross-talks, in
particular the substantial conversion from linear 
to circular polarization by the inclined mirrors of an image
derotator. Therefore, a derotator can strongly reduce the measurable
linear polarization depending on its orientation (\cite{deBoer20}),
which must be corrected with detailed hardware calibrations.
Also the Nasmyth telescope
mirror M3 requires polarimetric corrections (e.g. \cite{Joos08}),
which must be derived or at least verified with standard star data. 
Extremely helpful for the polarimetric calibration of many data sets
are objects with an unpolarized central star which serves as zero
polarization reference source.

For SPHERE-IRDIS there are moderate telescope and strong derotator
polarization effects which can be corrected with a data reduction
package based on a detailed polarization model of the whole instrument
(\cite{vanHolstein20}).
SPHERE-ZIMPOL uses three additional HWPs and one polarization
compensator plate, which actively correct for the strong telescope
polarization and derotator effects (\cite{Bazzon12}).

\begin{figure}[t]
\begin{center}
  \includegraphics[width=11.5cm]{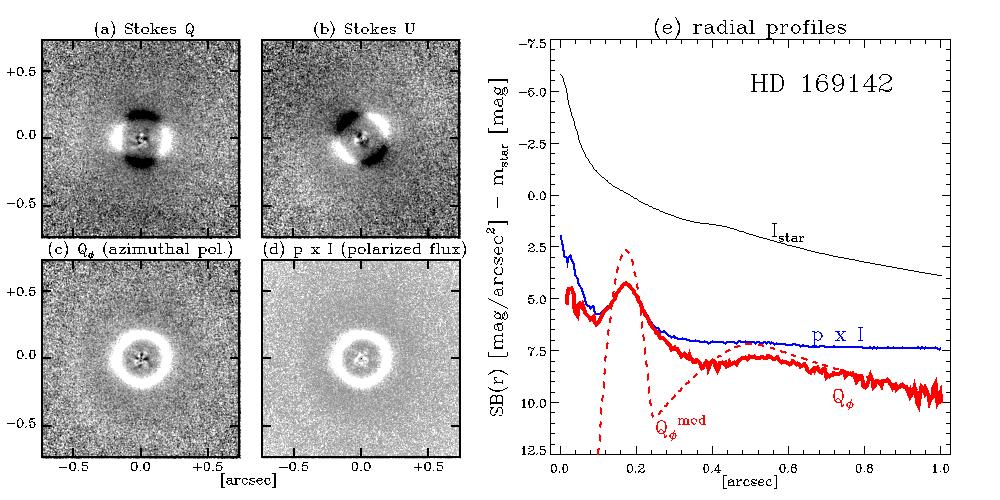}
\vspace*{-0.1 cm}
\caption{Images for (a) $Q$, (b) $U$, (c) $Q_\phi$, and (d) $p\times I$ 
  for the pole-on disk HD~169142 with a bright inner disk
  wall at $r=0.17''$ and an outer low surface brightness disk. Panel (e)
  gives azimuthally averaged radial profiles for $Q_\phi$, $p\times I$,
  the stellar PSF $I_{\rm star}$, and a model for the intrinsic $Q_\phi^{\rm mod}$
  (\cite{Tschudi21}), which fits after convolution the observed $Q_\phi$ curve.}
   \label{FigDemo}
\end{center}
\end{figure}

\subsection{Quantitative polarimetric measurements}
Often only differential polarization images
$Q_{\rm cs}(\alpha,\delta)$ and $U_{\rm cs}(\alpha,\delta)$ of a
circumstellar scattering source are available because the intensity
$I_{\rm cs}$ cannot be disentangled from the bright and highly
variable PSF of the central star.
The quantitative analysis of $Q_{\rm cs}$ and $U_{\rm cs}$ must
take noise bias effects and the polarimetric cancellation
into account.

If the $Q_{\rm cs}$ and $U_{\rm cs}$ images are noisy then
the determination of the polarized flux according to
$p_{\rm cs}\times I_{\rm cs}=(Q_{\rm cs}^2+U_{\rm cs}^2)^{1/2}$ is strongly
biased upward (\cite{Simmons85}) as can be seen
in Fig.~\ref{FigDemo} for the outer disk. This problem can be
avoided with the use of the azimuthal polarization parameters
$Q_\phi, U_\phi$. For circumstellar scattering the polarization
is essentially in azimuthal direction described by $Q_\phi$ and
the $U_\phi$-signal is very small. Therefore, the
$Q_\phi$-signal is a very good approximation of the polarized
flux $Q_\phi \approx p_{\rm cs}\times I_{\rm cs}$ and this parameter
avoids the noise bias (\cite{Schmid06a}).

The measurable polarization $Q_{\rm cs}, U_{\rm cs}$
is also affected by the limited observational resolution, 
which introduces smearing and for polarimetry also a 
cancellation between image regions with $+Q$ and $-Q$ or $+U$ and $-U$
signals (\cite{Schmid06a}). This effect can be very substantial for
AO polarimetry of compact sources as shown in Fig.~\ref{FigDemo}(e) for
the inner ring of HD~169142. There, the measurable signal $Q_\phi$ is
about 4 times smaller than the model for the intrinsic
signal $Q_\phi^{\rm mod}$ before convolution with the PSF. For the
more extended disk the cancellation effect is much reduced. This
effect depends strongly on the atmospheric conditions, but can be modeled
and corrected well, if the PSF is known accurately (\cite{Tschudi21}).

\section{Scientific results}

\subsection{Protoplanetary disks}

Protoplanetary disk are
the ``easy targets'' for polarimetric imaging with AO systems
especially the so-called transition disks. They have
a large central cavity and the strongly
illuminated inner disk wall can be separated by $d>10~$AU from the star
or $>0.1''$ for a disk in a nearby ($d\approx 100$~pc) star forming region.
The disk integrated
polarized flux is up to $Q_\phi/I_{\rm star} \approx 1~\%$.

The images show the scattering polarization of the dust
in the surface layer of optically thick disks and they reveal
a surprising diversity of hydrodynamical structures:
central disk cavities (\cite{Quanz11,Hashimoto12}),  
large scale spirals (\cite{Garufi13,Benisty15}, Fig.~\ref{FigScience}),
circular gaps (\cite{Quanz13}, Fig.~\ref{FigDemo}),
shadows from tilted disks near the star
(\cite{Marino15,Pinilla18}), 
and much more. Some structures are explained by the presence
of newly formed planets and two young planets were indeed found
in the large disk cavity of PDS~70 (\cite{Keppler18}). These
images and corresponding ALMA maps are of much interest
for the understanding of the planet formation processes in disks. 

Well calibrated disk data provide
the spectral dependence of the polarized reflectivity
$Q_{\rm disk}(\lambda)/I_{\rm star}(\lambda)$ and the few existing
measurements indicate a ``reddish'' color (e.g. \cite{Hunziker21}).
The fractional polarization $p_{\rm disk}$ of the reflected light 
was first measured with HST for AB Aur (\cite{Perrin09}) but
this can now also be achieved with AO systems. Reported
values for a scattering angle of $\approx 90^\circ$ range
from $p_{\rm disk}\approx 20-30~\%$ for visual wavelengths,
to higher values $p_{\rm disk}\approx 30-60~\%$ in the near-IR
(\cite{Monnier19,Hunziker21,Tschudi21}). More such quantitative
measurements will become available and they will help to constrain
the properties of the scattering dust.

\subsection{Debris disks}
Dust debris disk are signposts for the presence of colliding
planetesimals around a star. Typically, we see the scattered
light from dust rings at separations of $\approx 10-100$~AU,
sometimes surrounded by scattering halos from small grains blown
out by radiation pressure. Debris disk are optically thin, with
a low polarization
contrast $Q_\phi/I_{\star}\lapprox 0.05~\%$ (\cite{Engler17}) 
requiring deep observations with extreme AO systems like GPI or SPHERE.
Edge-on systems are easier to detect because the large grains are strongly
forward scattering (\cite{Esposito20}). 

\begin{figure}[t]
\begin{center}
  \includegraphics[width=8cm]{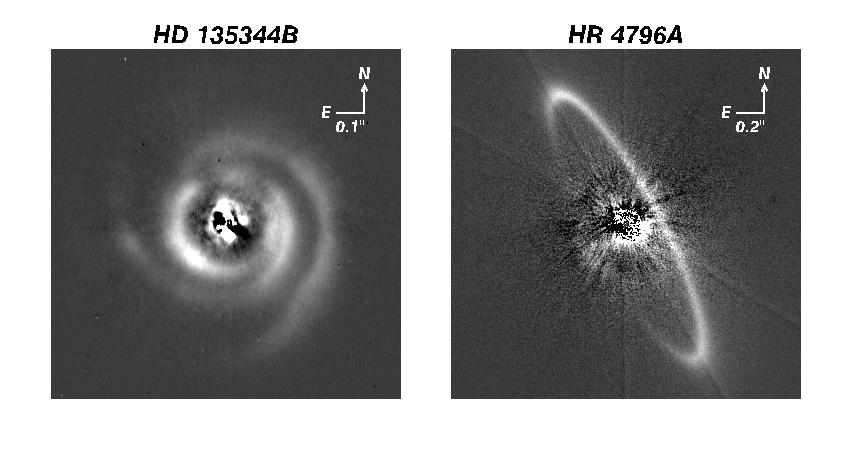}
\vspace*{-0.6cm}
\caption{$Q_\phi$-images of (a) the protoplanetary spiral disk HD~135334B 
  and (b) the bright debris disk HR~4796A. The data are described in
  Stolker et al.~(2016) and Milli et al.~(2019).}
   \label{FigScience}
\end{center}
\end{figure}

Debris disks with high inclination, like the bright system HR~4796A
(Fig.~\ref{FigScience}),
are ideal for the determination of the dust scattering phase function 
for the intensity and the polarization because the photons undergo
only one scattering and the scattering angles are well defined
(\cite{Perrin15,Milli19,Chen20,Arriaga20}).
Such data provide important information for model calculation of
the light scattering by complex dust particles.

\subsection{Dust formation in the stellar wind of red giants}

Important processes for the mass loss from red giants are the
formation of dust and the wind acceleration by the radiation
pressure acting on dust grains. The light scattered by the
dust can be directly observed with non-coronagraphic
AO polarimetry. The key technical requirement is a high spatial
resolution because the stellar diameters are only about 50~mas
for the most extended red giants
(e.g. $\alpha$ Ori, Mira, W Hya, R Dor, $\alpha$ Sco).
This favors strongly observations at small $\lambda$ with SPHERE-ZIMPOL
(Tab.~\ref{Systems}), which achieves a spatial resolution of
(${\cal{R}}\approx 20$~mas). Interesting monitoring results are
also obtained with an ``AO free'' system based on differential
speckle polarimetry (\cite{Safonov20}) and very promising are the
prospects for polarimetry using sparse aperture masking interferometry,
which will achieve a spatial resolution of up to 10~mas with the
new Subaru-SCeXAO-Vampires instrument (\cite{Norris15}).

The polarimetric observations show clumpy and variable dust structures,
and the derived radial distribution defines the locations for
dust formation and wind acceleration. From the color of the
scattered light one can estimate the dust particle sizes
(e.g. \cite{Ohnaka16,Khouri20}).

\subsection{Reflecting extrasolar planets}

Light reflecting from planets is polarized and this offers
the possibility for detecting extra-solar planets with high resolution
polarimetry. The SPHERE-ZIMPOL instrument
was optimized for this science case (\cite{Schmid06b})
and a search for extra-solar planets around the nearest bright stars
$\alpha$ Cen A and B, Sirius, $\epsilon$ Eri, Altair and $\tau$ Cet
has been carried out by Hunziker et al.~(2020). The requirements are extreme,
for example a Jupiter-sized planet at a separation of 1~AU would produce
only a polarimetric contrast of about
$C_{\rm pol}=p_{\rm planet}\times I_{\rm planet}/I_{\rm star}\approx 2\cdot 10^{-8}$
(e.g. \cite{Buenzli09}). Such detection limit could be reached
with an exposure time of $t_{\rm exp}=3.4^h$ for $\alpha$ Cen A but
no successful detection has been achieved. However, it is
shown that the ZIMPOL polarimetry is for $r>0.5''$ photon noise limited
and the contrast limit should just improve with $t_{\rm exp}$ like
$C_{\rm limit}\propto (t_{\rm exp})^{-1/2}$. Deeper searches for promising
targets are therefore ongoing for 
$\epsilon$ Eri b 
or the recently reported planet candidate around
$\alpha$ Cen A (\cite{Wagner21}).

\section{Conclusions} 

High resolution polarimetric imaging with AO systems made
in the last decade a major step forward from a 
special technique used by a few experts to a main stream
observing mode offered by leading observatories to the
astronomical community. Many astronomers use now this
technique and they produced a continuous string of first class
science results on circumstellar disks, dusty shells around
evolved stars, and several other science topics not covered
in this review.

AO polarimetry is not a simple observing technique, because 
of instrument polarization effects and the variable AO performance
which depends on the atmospheric turbulence. In addition, many
of the most interesting targets are very faint
and very close to the bright central star. Despite these extreme
challenges the results obtained with different systems
agree very well, giving much credibility to this technique and
proving its maturity. 

This success will lead to more progress in the
near and more distant future, like deeper observations for
fainter targets, improved observing and calibration procedures
for quantitative measurements, but also upgrades on existing instruments
and hopefully the development of much more powerful AO polarimeters
for the coming generation of extremely large telescopes.

\end{document}